\pdfoutput=1
\documentclass[12pt]{article}
\usepackage{graphicx}

\begin{document}
\title{Modeling a quantum Hall system via elliptic equations}
\author{Artur Sowa \\
Department of Mathematics and Statistics\\
University of Saskatchewan \\
106 Wiggins Road,
Saskatoon, SK S7N 5E6 \\
Canada \\
sowa@math.usask.ca $\qquad$  a.sowa@mesoscopia.com}
\date{}
\maketitle

\begin{abstract}
\noindent Quantum Hall systems are a suitable theme for a case study
in the general area of nanotechnology. In particular, it is a good
framework in which to search for universal principles relevant to
nanosystem modeling, and nanosystem-specific signal processing.
Recently, we have been able to construct a PDE model of a quantum
Hall system, which consists of the Schr\"odinger equation
supplemented with a special type nonlinear feedback loop. This
result stems from a novel theoretical approach, which in particular
brings to the fore the notion of quantum information. In this
article we undertake to modify the original model  by substituting
the dynamics based on the Dirac operator. This leads to a model that
consists of a system of three nonlinearly coupled first order
elliptic equations in the plane.
\end{abstract}

\section{Quantum entanglement of composite systems and its associated Hamiltonian dynamics}

In this section we give a very brief overview of the principles
behind the mesoscopic loop models as developed in
\cite{sowa5,sowa6,sowa7,sowa8, sowa9}. The theme of quantum
information is extremely relevant to our approach, but it has  been
brought to light in explicit terms only recently; an in-depth
discussion will appear elsewhere, \cite{sowa10}.

Since the measurement of Hall resistance in a \emph{quantum}-Hall
system results in a classical signal, it is natural to ask if any
quantum information is in the process transduced to the classical
medium. Let us develop this point of view into a formal discussion.
Suppose that the electronic solid-state system and the ambient
magnetic field are represented by two Hilbert spaces, say,
$\mbox{\textbf{H}}$ and $\widehat{\mbox{\textbf{H}}}$ respectively.
According to the principles of quantum mechanics, \cite{Sten}, the
combined system is then represented by the tensor product space
$\widehat{\mbox{\textbf{H}}} \bigotimes \mbox{\textbf{H}}$.
Moreover, the states of the composite system assume the form
\begin{equation}
 | \Psi _{\mbox{combined}}\rangle = \sum\limits_{m,n} A_{m n}^*|\varphi _m
\rangle \otimes |\psi_n\rangle \in \widehat{\mbox{\textbf{H}}}
\bigotimes \mbox{\textbf{H}}. \label{psicomb}
\end{equation}
We will associate with every state as above an \emph{entanglement
operator} (with no intention to imply that the composite state
(\ref{psicomb}) is necessarily entangled in the rigorous sense of
the term, \cite{Sten})
\begin{equation}
K = \sum A_{m n} |\varphi _m \rangle\langle\psi _n|.
\end{equation}
Thus, if an electronic system is found in the state $\psi\in
\mbox{\textbf{H}}$, then the electromagnetic system will collapse
onto the state
\[
K\psi \in \widehat{\mbox{\textbf{H}}}.
\]
Let us define the density operator of system $\mbox{\textbf{H}}$ in
the usual way
\begin{equation}
 \rho = \mbox{Tr}_{\widehat{\mbox{\textbf{H}}}} \left(| \Psi
_{\mbox{combined}}\rangle \langle \Psi _{\mbox{combined}}| \right).
\label{subrho1}
\end{equation}
One readily finds
\begin{equation}
 \rho = K^*K. \label{rhodecomp}
\end{equation}
This lays out the vocabulary for a quantum-mechanical description of
a composite system. Next, we postulate a model relevant specifically
to the quantum Hall systems. Namely, we propose that in the presence
of a magnetic field, the dynamics of the electronic system is
derived from the energy functional of the form
\[
\mbox{Tr} \left( H \rho \right) + \beta\log \det\left(\rho \right),
\]
where $H$ is the ``regular" Hamiltonian governing the microscopic
dynamics of the electronic subsystem. For the purposes at hand there
is no need to discuss the constant $\beta$ beyond the simple
statement that it is nonzero when the magnetic field is
nonvanishing. Furthermore, we assume for simplicity that $H$ is
diagonalizable in the basis consisting of its eigenvalues
\begin{equation}
H\psi _k = E_k \psi _k. \label{Heigenval}
\end{equation}
 Now, substituting the decomposition (\ref{rhodecomp}) into the
above yields the following Hamiltonian
\begin{equation}
\Xi (K) = \mbox{Tr} \left(K H K^* \right) + \beta\log \det\left(K^*K
\right). \label{Ksi}
\end{equation}
This in turn leads to the following system of Hamiltonian equations,
see \cite{sowa5} and \cite{sowa6}:
\begin{equation}
-i\hbar\dot{K} = KH  + \beta K^{*-1}.
 \label{mesoeq}
\end{equation}
(From this point on we will set  $\hbar =1$.) The corresponding
eigenvalue problem assumes the form
\begin{equation}
 KH  + \beta K^{*-1} = \nu K.
 \label{mesoeig}
\end{equation}
We postulate that homeostasis of the quantum Hall system is captured
by the latter equation. Solutions of (\ref{mesoeig}) are found to
assume the form (\cite{sowa5}, \cite{sowa6}):
\begin{equation}
K= \sum\limits_{E_k < \nu} \left(\frac{\beta}{\nu -
E_k}\right)^\frac{1}{2} |U\psi _k\rangle\langle \psi _k |.
\label{psi2K}
\end{equation}
where $U: \mbox{\textbf{H}}\rightarrow \widehat{\mbox{\textbf{H}}}$
is an arbitrary unitary transformation.

\section{The mesoscopic-loop model based on the Dirac-type equations}
The model we are about to present deals with two types of particles,
which will be called electrons and holes. Both of them will have the
same effective mass $m^* =1$, and charge $e=1$. It seems appropriate
to emphasize that, of course, the notion of \emph{effective} mass
depends on the model. All analysis will be performed in a
two-dimensional plane. Recall that the Hodge star $*$ in two
dimensions is a linear operation on differential forms determined by
the relations $*dx=dy,
*dy = -dx,
*1=dx\wedge dy, *(dx\wedge dy) =1$. The exterior derivative $d$ defines
 the co-derivative $\delta = -*d*$.
Let the vector potential $A$ be given in local coordinates as $A
=A_1 dx+A_2 dy$. In particular, $\delta A= - (A_{1,x}+A_{2,y})$,
while
\begin{equation} B (x,y)= *dA = A_{2,x}-A_{1,y}. \label{BdA}
\end{equation}
is the magnetic flux density. The correspondng Schr\"odinger
operator $S_A$ acts on a wavefunction $\psi$ as follows
\begin{equation}
2S_A\psi =  -(\partial ^2_x +\partial ^2_y)\psi + 2i
(A_1\partial_x\psi +A_2\partial_y\psi )+(A_1^2+A_2^2)\psi -i (\delta
A) \psi \label{Schrod}
\end{equation}
(The factor of $2$ in front of $S_A$ stems from the fact that in the
adopted units the kinetic energy is multiplied by a factor of $\hbar
^2/2m^*=1/2$.) The Schr\"odinger operator can be successfully used
in a construction of mesoscopic loop models, \cite{sowa9}. In this
article we will consider a modification  based on a Dirac type
factorization of the Schr\"odinger operator. To this end we use the
two-dimensional Dirac operator, \cite{Thaller}, with its parameters
set at $m=0$ and $c=1$. Namely, let
\begin{equation}
H_A = \left[\begin{array}{lr}
 0 & D^*\\
 D & 0\\
\end{array} \right],
\label{Dir_Hamilt}
\end{equation}
where
\begin{equation}
D = -i\partial_x -A_1 +\partial _y -iA_2, \label{D}
\end{equation}
so that
\begin{equation}
D^* = -i\partial_x -A_1 -\partial _y +iA_2 \label{Dstar}.
\end{equation}
A direct calculation shows that
\begin{equation}
H_A^2 = \left[\begin{array}{cc}
 D^*D  & 0\\
 0& DD^*\\
\end{array} \right]
= \left[\begin{array}{cc}
 2S_A - B  & 0\\
 0& 2S_A+B \\
\end{array} \right],
 \label{Dir_Hamiltsq}
\end{equation}
where $B=B(x,y)$ is determined by $A$ via (\ref{BdA}). Observe the
difference of signs at the magnetic flux density term. This
necessitates the interpretation of the two wavefunctions defining
the statevector
\[
\left|\begin{array}{c} \psi _1 \\ \psi_2 \end{array} \right>
\]
as representing particles with opposite \emph{pseudo-spins}. At this
point we venture to describe a quantum Hall system in a manner
discussed in the previous section, which brings the notion of
entanglement to the fore. Namely, we consider the following system
of equations
\begin{equation}
\begin{array}{rcl}
H_A \left|\begin{array}{c} \psi _1 \\ \psi_2 \end{array} \right> &=&
E \left|\begin{array}{c} \psi _1 \\ \psi_2 \end{array} \right>
\\
\\
 **dA (x,y) &=&
|K \left|\begin{array}{c} \psi _1 \\ \psi_2
\end{array}\right>(x,y)|^2,
\end{array}
\label{MeM_Dirac}
\end{equation}
which involves a coupling realized via the entanglement transform
$K$. Note that this forces an appropriate normalization of $K$, so
as to guarantee the correct charge to flux ratio.

\section{Quantization of Hall resistance}
Note that if $\left|\begin{array}{c} \psi _{1} \\
\psi_{2}
\end{array}\right>$ is a solution of (\ref{MeM_Dirac}), then in
particular it is an eigenfunction of  $H_A$. We now request that $K$
be a stationary solution as in (\ref{psi2K}) with $U=I$, where
\[|\psi_k \rangle = \left|\begin{array}{c} \psi _{k,1} \\ \psi_{k,2}
\end{array}\right>
\]
 are the eigenvectors of the Dirac
Hamiltonian $H_A$. Assume $\nu >E$. Without loss of generality we
may assume
that $\left|\begin{array}{c} \psi _{1} \\
\psi_{2}
\end{array}\right>$ is on the list of mutually orthogonal eigenfunctions defining
$K$ via (\ref{psi2K}). (This can always be arranged, even if $E$ has
multiplicity greater than $1$.) Thus, we have
\[
K \left|\begin{array}{c} \psi _1 \\ \psi_2
\end{array}\right> = c \left|\begin{array}{c} \psi _1 \\ \psi_2
\end{array}\right>,
\]
for a constant $c$. Consequently
\[
|K \left|\begin{array}{c} \psi _1 \\ \psi_2
\end{array}\right>(x,y)|^2 \sim |\psi _1|^2 +|\psi _2|^2.
\]
This implies that the last equation of (\ref{MeM_Dirac}) is
equivalent to the statement
\begin{equation}
\partial _x A_2 - \partial _y A_1 =R_H\left(|\psi _1|^2 +|\psi _2|^2\right).
\label{cyl_adaptx}
\end{equation}
Note that $R_H$ is the ratio of the total flux to the total charge
counted between the two types of particles. In particular, due to
quantization of flux and charge, $R_H$ is a rational number when
expressed in appropriate units. We will demonstrate that in a
symmetric realization of this model the Hall resistance is equal to
$2R_H$. We conduct the analysis in a rectangular-domain setting, and
with the assumption that the wavefunctions are periodic in $x$, both
with the same period, i.e.
\[
\begin{array}{c} \psi _1 = \exp{(ikx)}\chi_1(y) \\ \psi _2 = \exp{(ikx)}\chi_2(y). \end{array}
\]
By substituting to (\ref{MeM_Dirac}), we readily obtain
\begin{equation}
\begin{array}{lll}
\chi_1' &= &-(k-A_1)\chi_1 +E \chi_2 \\
\chi_2' &= &(k-A_1)\chi_2 -E \chi_1 .
\end{array}
\label{cyl_chi}
\end{equation}
Furthermore, let us assume $A_2=0$, and $A_1=A_1(y)$, $A_1(0)=0$.
Let us also introduce an auxiliary variable $w=k-A_1$. In such a
case (\ref{MeM_Dirac}) is equivalent to the system
\begin{equation}
\begin{array}{lll}
w' &= & R_H (\chi_1^2 +\chi_2^2),\quad w(0)=k\\
\chi_1' &= &-w\chi_1 +E \chi_2 \\
\chi_2' &= &w\chi_2 -E \chi_1 .
\end{array}
\label{cyl_chiw}
\end{equation}
By differentiating  we readily obtain
\begin{equation}
\begin{array}{lll}
\chi_1'' &= & (-E^2 +w^2-w') \chi_1 \\
\chi_2'' &= & (-E^2 +w^2+w') \chi_2 .
\end{array}
\label{cyl_chiSch}
\end{equation}
These are $1$-D stationary Schr\"odinger equations with energy
$E^2/2$. Note that $\chi_1$  is affected by potential $(w^2-w')/2$,
while $\chi_2$ is affected by potential $(w^2+w')/2$. Both
quantities result from the presence of the magnetic field, and thus
represent the Hall effect.  There are two potential functions,
differing by the sign of the correction term $w'$, because two types
of carriers are involved in the charge transport.  Note that
\begin{equation}
\label{wexplic} w(y) = k+R_H\int\limits_0^y (\chi_1^2(y')
+\chi_2^2(y'))dy'.
\end{equation}
We calculate the current next. Recall that
\[
\label{currentrel} j= Re\{\psi_1 ^* (i \cdot d \psi_1 + \psi_1 A)\}
+ Re\{\psi_2 ^* (i \cdot d \psi_2 + \psi_2 A)\}. \] A direct
calculation shows that the two $1$-forms amount to
\[
 Re\{\psi_l ^* (i \cdot d \psi_l + \psi_l A)\} = -\chi_l^2(y) w(y)
dx,\quad l=1,2.
\]
In summary
\begin{equation}
\label{Hallcurrel} j=-\frac{1}{R_H}w'(y)w(y)dx=
\frac{1}{R_H}*d\frac{w^2}{2}.
\end{equation}
(We remind the reader that $*$ denotes the Hodge star.) Observe that
the current flows along the $x$-axis. Since both types of carriers
contribute to the current, the total Hall potential is the sum of
the two separate potentials pointed out above, i.e.
\begin{equation}
\label{Hallpot} V_H = (w^2+w')/2 + (w^2-w')/2 = w^2.
\end{equation}
It follows that
\begin{equation}
\label{HallR} j= \frac{1}{2R_H}*dV_H.
\end{equation}
 By integrating we then obtain that the
Hall resistance is $2R_H$. We remark that this is twice the value
obtained in the Schr\"odinger type model, cf. \cite{sowa8},
\cite{sowa9}. Recall that $R_H$ has been introduced as the ratio of
the number of quanta of the magnetic field to the number of quanta
of electric charge $e=1$ of both types of particles. If particles of
type, say, $\psi _2$ were not accounted for in the calculation of
the filling factor, then there would be a discrepancy between the
inverse filling factor and the quantity $R_H$.

It is interesting to ask when the electrons and holes are engaged as
charge carriers in equal numbers. Let us position the edges of the
plate symmetrically about the x-axis, i.e. $y\in [-b,b]$. Note that
the number of carriers of each type is proportional to
\begin{equation}
\int\limits_{-b}^b \chi_l^2(y')dy', \quad l=1,2. \label{density}
\end{equation}
 Now assume $\chi_2(0)=\chi_1(0)$, and consider the last two equations of
(\ref{cyl_chiw}). In such a case one can find a solution satisfying
the condition $\chi_2(y)=\chi_1(-y)$. For this solution particles of
each type occur in equal numbers.

\section{The elliptic system in the plane}
In the previous section we demonstrated that Hall resistance is
quantized under certain symmetry assumptions, which reduce the
problem to a system of ODEs. We emphasize that Hall potentials and
Hall resistance can be defined and discussed in the general
two-dimensional setting, \cite{sowa9}. In fact, the two-dimensional,
i.e. PDE, setting cannot be avoided if we want to study the effect
of lattice potential, etc. We will now briefly discuss the
ramifications of the PDE problem. Consider (\ref{MeM_Dirac}) jointly
with the assumptions about $K$ put forth in the previous section. In
this case it is equivalent to the following system of first order
partial differential equations
\begin{equation}
\begin{array}{lll}
\partial _y \psi_1 &= &i \partial _x\psi_1 +(A_1+iA_2)\psi_1 +E \psi_2 \\
\partial _y \psi_2 &= &-i \partial _x\psi_2 -(A_1-iA_2)\psi_2 -E\psi_1 \\
\partial _y A_1 &= &\partial _x A_2 -R_H\left(|\psi _1|^2 +|\psi _2|^2\right).
\end{array}
\label{cyl_adapt}
\end{equation}
Note that the system will be strictly elliptic, if it is amended
with a gauge condition
\begin{equation}
\partial _y A_2 = -\partial _x A_1.
\label{gauge}
\end{equation}
Note that the nonlinear terms in (\ref{cyl_adapt}) are (real)
analytic functions of the dependent variables. Therefore the
classical Cauchy-Kovalevskaya Theorem, \cite{Cour-Hilb}, guaranties
local existence of solutions of the initial value problem when the
boundary conditions are also analytic. The initial value problem may
help gain some insight into the nature of the system; an example of
a solution obtained numerically with a finite difference method is
displayed in Fig. 1. 
Since the boundary value problems spring to mind in this context,
one should emphasize that in view of ellipticity, the natural
setting is that of the Riemann-Hilbert type problem,
\cite{Wendland}. 
However, we wish to signal briefly that the boundary value problem
is not the only type of approach appropriate for the study of the
mesoscopic-loop model; there is a promising complementary approach.
Namely, setting $A=A_1+iA_2$, we can represent equations
(\ref{cyl_adapt})-(\ref{gauge}) as a system of three nonlinearly
coupled $\overline\partial$-equations:
\begin{equation}
\begin{array}{lll}
\bar{\partial} \psi_1 &= &\frac{i}{2}( A\psi_1 +E \psi_2) \\
\bar{\partial} \bar{\psi_2} &= &-\frac{i}{2} (A\bar{\psi_2} +E\bar{\psi_1}) \\
\bar{\partial} \bar{A} &= &-\frac{i}{2} R_H(|\psi _1|^2 +|\psi
_2|^2).
\end{array}
\label{delbar}
\end{equation}
This suggests the relevance of a variety of complex methods,
\cite{Krantz}, but this theme will not be discussed in the present
article. \vspace{.5cm}

\noindent \textbf{Closing comments:} The Dirac equation is
undergoing a renaissance in condensed matter physics due to its
relevance to the unusual two-dimensional crystal called graphene,
e.g. \cite{Novoselov3}. I emphasize that in this article we have not
addressed any of the issues related to that stream of research.
Rather, we have used the Dirac equation as an analytic alternative
to the Schr\"odiner equation. It helped us find an almost equivalent
reformulation of the nonrelativistic model.

\newpage

\begin{figure}[h]
\centerline{ \includegraphics[height=75mm]{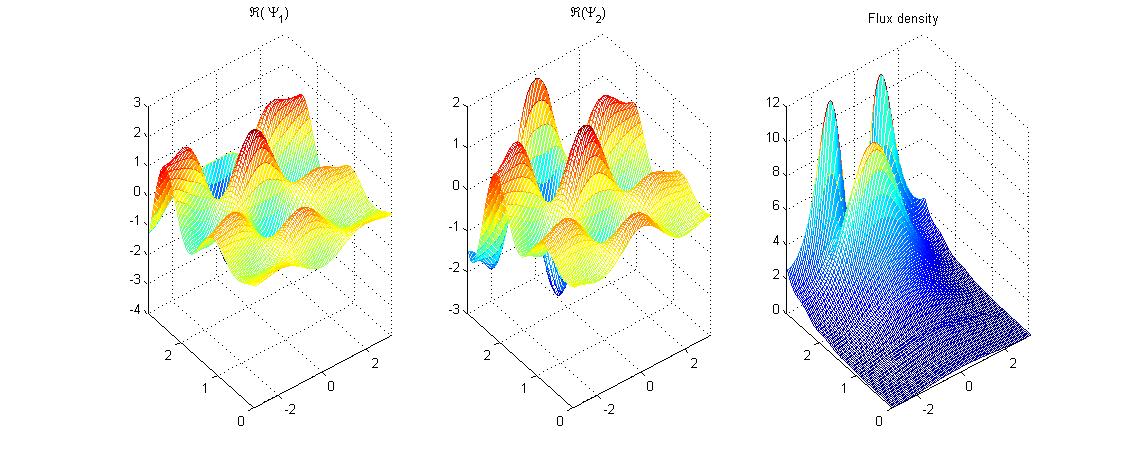}} \caption{A numerical solution of
(\ref{cyl_adapt}) with $E=5$ (in units adopted throughout the
article), $R_H = \frac{3}{5}$, and $A_2 \equiv 0$. The
 initial values $\psi_2(x,0) = 0.3\exp(ix) +0.2*\exp (2ix)$, $\psi_1(x,0) =-0.9\psi_2(x,0)$,
 and $A_1(x,0) =0$ are prescribed on the edge facing
front-right.}
\end{figure}

\end{document}